# Strongly Correlated Polaritons in a Two-Dimensional Array of Photonic Crystal Microcavities

Neil Na,[1] Shoko Utsunomiya,[2] Lin Tian,[1] and Yoshihisa Yamamoto[1,2]

[1]*E. L. Ginzton Laboratory, Stanford University, Stanford, CA 94305, USA*

[2]*National Institute of Informatics, Hitotsubashi, Chiyoda-ku, Tokyo 101-8430, Japan*

We propose a practical scheme to observe the polaritonic quantum phase transition (QPT) from the superfluid (SF) to Bose-glass (BG) to Mott-insulator (MI) states. The system consists of a two-dimensional array of photonic crystal microcavities doped with substitutional donor/acceptor impurities. Using realistic parameters, we show that such strongly correlated polaritonic systems can be constructed using the state-of-art semiconductor technology.

Quantum many-body systems, such as strongly correlated electrons, are generally difficult to understand due to the lack of appropriate theoretical tools. A brute-force matrix diagonalization method is limited by the exponentially growing Hilbert space with the number of particles. A quantum Monte-Carlo simulation method often suffers from the so-called sign problem. An analytical mean field method provides good approximate solutions for three-dimensional systems but limited applications for two-dimensional systems. An interesting alternative is to construct a quantum simulator that implements a model Hamiltonian with controllable parameters [1]. One such example was demonstrated using Bose-condensed cold atoms in an optical lattice. The predicted QPT from the SF to MI states [2] was observed by changing the ratio of on-site repulsive interaction to hopping matrix element [3]. Recently, by using speckle lasers or multi-color optical lattices, the novel quantum phases of a disordered system such as BG have been explored both theoretically [4] and experimentally [5]. These results opened a door for simulating complex many-body systems with more controllable artificial systems.

Simulating the Bose-Hubbard model (BHM) using polaritons has recently attracted intense interest [6,7,8]. These schemes require a coupled high-$Q$ cavity array where either EIT in a four-level atomic ensemble or single-atom cavity QED in the strong coupling regime is employed. In this letter, we show that such strongly correlated polaritonic systems can be constructed using a simpler approach. Our scheme consists of a two-dimensional array of photonic crystal microcavities [9] doped with substitutional donor/acceptor impurities [10]. A schematic plot is shown in Fig. 1. The polaritons hop from site to site via optical field coupling between each microcavity, and interact with each other through the nonlinearity induced by the strong coupling of cavity photons and bound excitons. Using realistic experimental parameters, we show that our system undergoes a polaritonic QPT from the SF to MI states by calculating the photon order parameter. Losses due to cavity photon leakage and bound exciton decay are taken into consideration to evaluate the equilibrium condition. The formation of BG states due to the inevitable experimental disorders such as fluctuations of cavity photon resonance frequency, photon-exciton coupling constant and cavity impurity number is examined. These imperfections degrade the possibility to observe a polaritonic MI but serve as natural implementation of polaritonic BG. The proposed scheme combines the large oscillator strength and small inhomogeneous linewidth of donor/acceptor-bound excitons embedded in bulk semiconductor matrix [11], and the recent advancement in photonic crystal microcavities with high cavity *Q* factor and small mode volume [12,13,14]. Moreover, our scheme is based on many-exciton cavity QED effect and therefore no need to control the cavity impurity number precisely one [7,8].

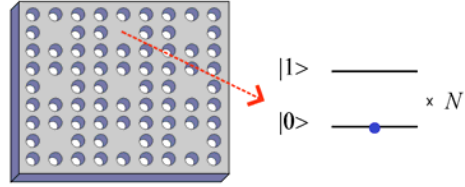

**FIG. 1 (color online). Schematic plot for a two-dimensional array of photonic crystal microcavities doped with substitutional donor/acceptor impurities. *N* impurities per cavity is assumed. Impurity-bound excitons are modeled as isolated two-level atoms.**

We start our analysis by considering the optical field coupling between adjacent microcavities. The tunneling part of the Hamiltonian is given as

$$H_{\text{tunneling}} = -t \sum_{<ij>} a_i^\dagger a_j \qquad (1)$$

where $<ij>$ indicates only the nearest neighbor coupling. *t* is the photon tunneling energy determined by the overlapping of the nearest neighbor cavity fields, and $a_i$ is the annihilation operator of the *i*th site cavity mode. To quantitatively estimate the condition of QPT, we perform a mean field analysis by applying the decoupling approximation [7,15] i.e. let $a_i^\dagger a_j = <a_i^\dagger> a_j + a_i^\dagger <a_j> - <a_i^\dagger><a_j>$ and define a real-valued photon order parameter $\psi = <a_i>$. (1) can then be rewritten as

$$H_{\text{tunneling}} = \sum_i H_{\text{tunneling}}^i = \sum_i \left\{ -zt\psi(a_i^\dagger + a_i) + zt\psi^2 \right\} \qquad (2)$$

where *z* is the number of nearest neighbors. Next we consider the free and interacting part of the total Hamiltonian

$$H_{\text{free}} = \sum_i H_{\text{free}}^i = \sum_i \left\{ \omega_{ph} a_i^\dagger a_i + \omega_{ex}(L_z^i + N/2) \right\}, \qquad (3)$$

$$H_{\text{interacting}} = \sum_i H^i_{\text{interacting}} = \sum_i \left\{ g(L^i_- a_i^\dagger + L^i_+ a_i) \right\}. \tag{4}$$

$\omega_{ph}$, $\omega_{ex}$ and $g$ are the cavity photon resonance frequency, bound exciton transition frequency and photon-exciton coupling constant. $N$ is the cavity impurity number. $L_z$ is the collective angular momentum operator in the $z$ direction, and $L_\pm$ are the collective creation/annihilation operators. For the time being, we assume $g$ is constant for different impurities and $N$ is a fixed integer for all cavities. The single site eigenenergy spectrum considering only the free and interacting Hamiltonian is sketched in Fig. 2. In general, the number of eigenstates for each excitation manifold $n$ is equal to $n+1$ if $n<N+1$ and equal to $N+1$ if $n>N$. The polariton ground state (the lower branch of the $n=1$ excitation manifold) interaction energy $U$ can then be identified by calculating the energy cost to inject a second quasi-particle into the cavity. Notice that in the small detuning regime, $U$ is roughly proportional to $1/N$ if $N$ is larger than $n$. Such tr end is due to the fact that a large collection of two-level atoms behaves linearly (the collective angular momentum operator satisfies a bosonic commutation relation $[L_+,L_-]\sim N$ [16]) so that the polaritons are boson-like if $N$ is large.

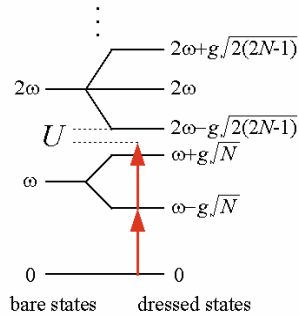

**FIG. 2 (color online). The eigenenergy spectrum of a many-exciton cavity QED system at zero frequency detuning. *N*>1 is assumed. The polariton ground state interaction energy *U* for adding a subsequent polariton to the cavity is labeled.**

Equipped with Hamiltonian (2), (3) and (4), we are now in position to evaluate the phase diagram. We consider only a single site Hamiltonian

$$H^i_{total} = H^i_{tunneling} + H^i_{free} + H^i_{interacting} - \mu(L^i_z + N/2 + a_i^\dagger a_i) \quad (5)$$

where $\mu$ is the chemical potential in grand canonical ensemble. Given $t$ and $\mu$, one can calculate the eigenenergies by diagonalizing (5) using bare states as a complete set of basis. The ground state energy can be found by minimizing the lowest eigenenergy with adjusting the photon order parameter. The accuracy of such calculation depends on how many state vectors are used to span the Hilbert space. The convergence of the eigenenergies usually indicates that a large enough basis set is considered. Here, the numerical value of optical wavelength is chosen to be 817 nm, which corresponds to the Si donor-bound exciton emission. $g$ is estimated as 33.3 GHz by calculating the bound exciton oscillator strength using the experimentally measured 1 ns lifetime, and a cavity mode volume equal to $(817/3.6)^3$ nm$^3$. 3.6 is the refractive index of GaAs. In Fig. 3, we plot the photon order parameter as a function of $t$ and $\mu$ given $N$=8, $\Delta=\omega_{ph}-\omega_{ex}$=0 and $z$=4. Notice that $N$=8 corresponds to a low bulk doping density 6.8×10$^{14}$ cm$^{-3}$ so the two-level atom approximation is still valid. Unlike the single-atom cavity QED systems [7,8], the Mott lobe sizes in the chemical potential direction are relatively unchanged for low filling factors. The nature of such a polaritonic QPT is neither purely fermionic nor bosonic, but shares similar features with the Bose-Hubbard model: localization of one additional particle per cavity upon entering the next Mott lobe.

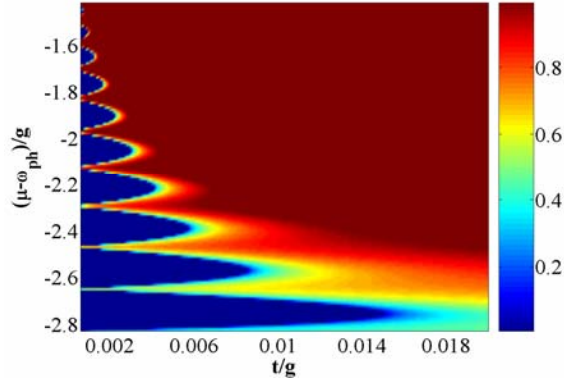

FIG. 3 (color online). Zero-temperature phase diagram obtained by plotting the photon order parameter as a function of $t$ and $\mu$. $N=8$, $\Delta=0$ and $z=4$ are assumed. The lowest Mott lobe corresponds to one polariton per cavity; the second lowest Mott lobe corresponds to two polaritons per cavity and so forth.

To understand the general behavior of the QPT condition, we plot the critical photon tunneling energy $t_c$ as a function of $N$ and $\Delta$ in Fig. 4a. The polariton ground state interaction energy $U$ is plotted in Fig. 4b. As previously explained, the system behaves linearly if $N$ is sufficiently large and hence both $t_c$ and $U$ diminish as $N$ increases. When using a blue detuning i.e. $\Delta>0$, both $t_c$ and $U$ are enhanced because the polaritons are more like their bound exciton components. The ratio of $U$ to $|c_{ph}|^2 t_c$ is plotted in Fig. 4c. $|c_{ph}|^2 t_c$ is the critical polariton tunneling energy where $|c_{ph}|^2$ is the photon fraction. As $N$ increases, these ratios approach 23.2, which is the predicted value for the infinite dimensional BHM [17]. Such trend suggests that our polaritonic system mimics the Bose-Hubbard dynamics when $N$ is sufficiently large. Nevertheless, from Fig. 4c, even with a few impurities in each cavity the system can still be reasonably described by the conventional BHM.

The polariton radiative loss arisen from the finite coupling to the environmental field reservoir provides useful information on the system dynamics. However, if the polariton loss rate is much faster than the polariton tunneling rate, the system never reaches an equilibrium state described by Hamiltonian (5). To observe the critical phenomenon in equilibrium, we arrive at the required cavity $Q$ factor as

$$Q_r \gg \frac{|c_{ph}|^2 \omega_p}{|c_{ph}|^2 t_c - |c_{ex}|^2 \dfrac{F}{\tau_e}}. \tag{6}$$

$\tau_e$ is the bound exciton spontaneous emission lifetime, and $F$ is the Purcell factor due to the inhibition of spontaneous emission in a photonic crystal [18,19]. We plot $Q_r$ in Fig. 4d as a function of $N$ and $\Delta$ where $F=0.2$ is used by setting the critical polariton tunneling rate to be 10 times larger than the polariton loss rate. In general, $Q_r$ about $10^6$ is needed. With a blue detuning, $Q_r$ can be relaxed by choosing an appropriately large blue detuning e.g. $Q_r \sim 10^5$ when $N=3$ and $\Delta=12g$. Such numbers are demanding but within the reach of state-of-art photonic crystal microcavity technology.

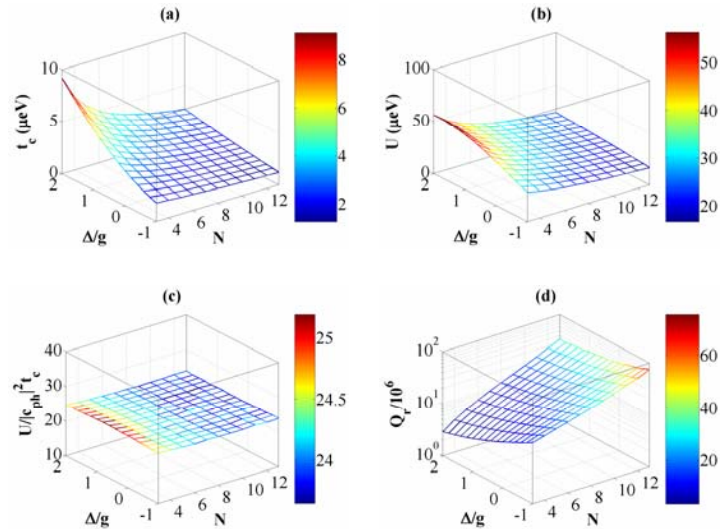

FIG. 4 (color online). System parameters plotted as a function of $N$ and $\Delta$: (a) critical photon tunneling energy (b) polariton ground state interaction energy (c) Condition of QPT from the SF to MI states (d) required cavity $Q$ factor.

We have shown that the many-exciton cavity QED Hamiltonian (5) generates similar dynamics to the conventional BHM, especially for the limiting case when $N$ is large. Another limiting case where the conventional BHM could be implemented is to use a red detuning that is much larger than the collective Rabi frequency. The photons and excitons are weakly coupled in this case, and the photon-like ($|c_{ph}|^2 \sim 1$) polariton ground state can be described by an effective Hamiltonian

$$H = \int_V \frac{1}{2}\varepsilon_o K_C \mathbf{E}^2 d\mathbf{r}^3 + \int_V \frac{1}{2}\mu_o \mathbf{H}^2 d\mathbf{r}^3 + \int_V \frac{1}{2}\varepsilon_o \chi_C^{(3)} \mathbf{E}^4 d\mathbf{r}^3 . \tag{7}$$

$\mathbf{E}$ and $\mathbf{H}$ are electric and magnetic field operators. $\varepsilon_o$ and $\mu_o$ are the free space permittivity and permeability. $K_C$ is the position dependent dielectric constant of the coupled photonic crystal microcavities and $\chi_C^{(3)}$ is the position dependent optical Kerr nonlinearity. All of the photon-exciton interactions can be renormalized into the linear and nonlinear dielectric constants by adiabatically eliminating the excitonic degrees of freedom. By substituting the electric and magnetic field operators into (7), which consists of the localized Wannier function based on isolated cavity field, we arrive at

$$H = -t\sum_{<ij>} a_i^\dagger a_j + \frac{U}{2}\sum_i a_i^\dagger a_i^\dagger a_i a_i \tag{8}$$

where

$$t = 2\varepsilon_o \int_V K_C(\mathbf{r})\varphi(\mathbf{r})\varphi(\mathbf{r}-\mathbf{d})d\mathbf{r}^3 , \tag{9}$$

$$U = -6\varepsilon_o \int_V \chi_C^{(3)}(\mathbf{r})\varphi^4(\mathbf{r})d\mathbf{r}^3 . \tag{10}$$

$\varphi$ is the cavity mode electric field assumed to be real and scalar, and **d** is the inter-cavity distance. Arriving at (8), we consider only the nearest neighbor photon hopping and on-site photon-photon interaction under rotating wave approximation. Notice that $U$ here is enhanced by increasing $N$, which is opposite to the case at near resonance, but can be understood by the linear proportionality between the optical Kerr nonlinearity and doping density [20]. $Q_r$ is about $10^8$ in order to operate the system in this limit, and therefore imposes great difficulty on experimental demonstration.

The idealized analysis presented above has neglected any possible system disorders. To further justify the experimental feasibility of our proposal, we consider the following impacts of system imperfection. First, $\omega_{ph}$ may fluctuate from sample to sample because of lithography error. Second, due to the inability to precisely position the impurities and the fact that the cavity field is spatially inhomogeneous, different $g_k$ for different impurity is expected. Finally, controlling a dose so that $N$ is a fixed integer for each cavity is not within the reach of the latest semiconductor fabrication technology. The BG states are naturally formed by taking the above parameter fluctuations i.e. $\Delta\omega_{ph}$, $\Delta g_k$ and $\Delta N$ into account. Unlike the MI states that are incompressible with gapped excitations, these states are compressible and gapless while still being insulating. Based on the strong coupling expansion in the thermodynamic limit [21,22], the Mott lobes on the phase diagram for each filling factor $n$ shrinks and eventually disappears when

$$2\Delta E + (2n-1)\Delta U \geq U . \qquad (11)$$

$\Delta E$ and $\Delta U$ stands for the ground state and interaction energy fluctuations from site to site respectively. In our system, $\Delta E$ and $\Delta U$ are complicated functions of $\Delta\omega_{ph}$, $\Delta g_k$ and $\Delta N$. To calculate their relation, we perform a statistical calculation by assigning a normal distribution to $\omega_{ph}$, a uniform distribution bounded from above by $g$ to $g_k$, and a Poisson/sub-Poisson distribution to $N$ [23]. We first examine whether an equilibrium polaritonic QPT from the BG to MI states exists when increasing the strength of these three disorders. As shown in Fig. 5, such information is obtained by plotting the iso-surface in which the critical polariton tunneling rate at the BG-MI phase boundary is set to be 10 times larger than the polariton loss rate, assuming $Q=10^6$, $<N>=3$ and $\Delta=12g$. Inside the iso-value surface, QPT from the SF to BG to MI states can be observed by continuously decreasing $t$. Outside the iso-value surface, only SF and BG phases exist. The maximal $\Delta\omega_{ph}$, $\Delta g_k$ and $\Delta N$ on the iso-surace are roughly 32 GHz, 0.14$g$ and 0.18$<N>$ respectively. Notice that the large tolerance of $\Delta\omega_{ph}$ is due to the fact that a large blue detuning is used. $\Delta g_k$ can be reduced by carefully designing the cavity field e.g. work on a monopole cavity mode. Controlling $\Delta N$ is relatively difficult but possible with the recent advances in few-ion implantation techniques.

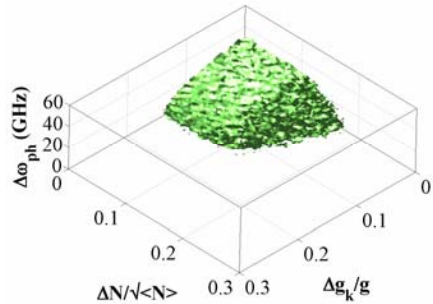

**FIG. 5 (color online). The iso-value surface that separates the existence of QPT from the BG to MI states as a function of $\Delta\omega_{ph}$, $\Delta g_k$ and $\Delta N$.**

Experimentally observing the polaritonic SF, BG and MI states requires the ability to change $t$ and $U$. $t$ can be easily controlled by designing the inter-cavity distance during the fabrication. $U$ can be controlled by adjusting the frequency detuning. For example, applying an electric field or varying the temperature can modulate $\omega_{ex}$ while keeping the $\omega_{ph}$ constant. Alternatively, injecting a molecular gas can modulate $\omega_{ph}$ while keeping $\omega_{ex}$ constant. By carefully designing the optical pump pulse, the system ground state starting with filling factor equal to zero and then adiabatically increased to one could be prepared. Then, the vertical cavity leakage serves as a natural probe on the signature of QPT. In the SF phase, the far field interference features a perfect visibility and lattice-like pattern. While undergoing QPT from the SF to BG states, such features degrade swiftly due to the lost of long-range coherence, which can be used to identify the phase boundary. In addition, by injecting photons via end-firing the membrane layer and measuring their transport through the whole structure, high and low transmissions distinguish the SF and BG phases. QPT from the BG to MI states can be characterized by spatially mapping the $g^{(2)}(\tau=0)$ measurements using a near field probe. For filling factor equal to one, $g^{(2)}(\tau=0)$ approaches zero when entering the MI phase for a equivalent large array of single photon sources are prepared. In addition, by spectrally measuring a transition from gapless to gapped excitations, the BG and MI states can be further confirmed.

In conclusion, we propose a practical scheme to observe the polaritonic QPT. Our scheme is based on two recent experimental breakthroughs, highly homogeneous substitutional donor/acceptor impurities in semiconductor, and photonic crystal microcavities with high cavity $Q$ factor and small mode volume. Our model belongs to an extended BHM, where the conventional BHM is recovered when the cavity impurity number is large or a large red frequency detuning is used. Novel QPT from the SF to BG to MI states can be prepared and measured using concurrent experimental quantum optics techniques. Finally, we point out that due to the flexibility of designing microcavity array topology, complicated systems such as supersolid [24] or the recently reported Bose Mott-glass [25] could also be simulated by our scheme.

Y.C. Neil Na is partially supported by MediaTek Fellowship. This work is partially supported by SORST program of Japan Science of Technology Corporation (JST) and NTT Basic Research Laboratories.